\documentclass[aps,prb,twocolumn,groupedaddress,showpacs]{revtex4-1}
\usepackage{color}
\usepackage{graphicx}
\usepackage{dcolumn}
\usepackage{bm}
\usepackage[mathlines]{lineno}
\usepackage[colorlinks,linkcolor=blue,anchorcolor=blue,citecolor=blue,urlcolor=blue,hyperindex,CJKbookmarks,dvipdfm]{hyperref}

\usepackage{amsmath}

\begin{document}

\title{Prediction of phonon-mediated superconductivity in borophene}

\author{Miao Gao$^{1}$}\email{gaomiao@nbu.edu.cn}
\author{Qi-Zhi Li$^{1}$}
\author{Xun-Wang Yan$^{2}$}
\author{Jun Wang$^{1}$}

\affiliation{$^{1}$Department of Microelectronics Science and Engineering, Faculty of Science, Ningbo University, Zhejiang 315211, P. R. China}

\affiliation{$^{2}$School of Physics and Electrical Engineering, Anyang Normal University, Henan 455000,
P. R. China}


\begin{abstract}
Superconductivity in two-dimensional compounds is widely concerned, not only due to its application in constructing nano-superconducting devices, but also for the general scientific interests.
Very recently, borophene (two-dimensional boron sheet) has been successfully grown on the Ag(111) surface, through direct evaporation of a pure
boron source. The experiment unveiled two types of borophene structures, namely $\beta_{12}$ and $\chi_3$.
Herein, we employed density-functional first-principles calculations to investigate the
electron-phonon coupling and superconductivity in both structures of borophene. The band structures of $\beta_{12}$ and $\chi_3$ borophenes
exhibit inherent metallicity. We found electron-phonon coupling constants in the two compounds are larger than that in MgB$_2$.
The superconducting transition temperatures were determined to be 18.7 K and 24.7 K through McMillian-Allen-Dynes formula.
These temperatures are much higher than theoretically predicted 8.1 K and experimentally observed 7.4 K superconductivity in graphene.
Our findings will enrich the nano-superconducting device applications and boron-related material science.
\end{abstract}


\maketitle

\section{Introduction}
The discovery of graphene provides a fascinating playground for nanoscale device applications in the future \cite{Geim-Nature6}. This has triggered a surge of interest for the synthesis and investigation of two-dimensional (2D) compounds,
such as silicene \cite{Aufray-APL,Vogt-PRL108,Du-ACSNano8}, germanene \cite{Cahangirov-PRL102,Bianco-ACSNano7,Derivaz-NanoLett15}, transition metal
dichalcogenides \cite{Geim-Nature499,Wang-Nature7,Radisavljevic-Nature6}, and phosphorene \cite{Das-ACSNano8,Li-Nature9,Qiao-Nature5}.
Borophene is another important 2D material which is also expected to act as a new functional compound or a precursor to construct boron nanotubes \cite{Tang-PRL99}.
2D triangular lattice, with buckling arrangement of boron atoms, is a metastable structure of boron sheet.
Theoretical studies suggested a set of more stable free-standing boron sheets by introducing some vacancy sites
in the triangular lattice. The predicted boron sheets take either planar or buckled geometries, depending on the
configurations of vacancies \cite{Yang-PRB77,Penev-NanoLett12,Yu-JPCC116,Tang-PRB82,Wu-ACSNano6}.

Very recently, the monolayer boron sheet has been successfully grown on the Ag(111) surface under ultrahigh vacuum \cite{Mannix-Science,Feng-arXiv1},
following the suggestion given by the first-principles simulation \cite{Liu-Angew52}.
The scanning tunneling microscopy images obtained in experiment revealed two distinct phases of boron sheet
for different temperatures of the substrate.
One is a striped phase, the other is a homogeneous phase.
Mannix and co-workers suggested the striped phase should be a buckled triangular lattice without vacancy \cite{Mannix-Science}. However, B. Feng \emph{et al.} attributed the striped phase to a $\beta_{12}$ model of boron sheet with vacancy \cite{Feng-arXiv1}.
More importantly, the
computed band structure based on a $\beta_{12}$ model of boron sheet with vacancies [see Fig.~\ref{fig:Structure}(a)],
is in excellent agreement with the observed angle resolved photoelectron spectroscopy \cite{Feng-arXiv2}.
A calculation, combining cluster expansion method with density-functional theory (DFT) found the ground-state structure of boron sheet on Ag(111) surface is $\beta_{12}$ \cite{Zhang-Angew127}. Another simulation about the nucleation of boron on Ag(111) surface again
prefer the $\beta_{12}$ model \cite{Xu-arXiv1601}.
On the other hand, the homogeneous phase of boron sheet corresponds to $\chi_3$ structure [see Fig.~\ref{fig:Structure}(b)] \cite{Feng-arXiv1}.
Hereafter, we named these two boron sheets as $\beta_{12}$ borophene and $\chi_3$ borophene, respectively.

One important application of 2D compounds is driving them to superconducting states to manufacture nano-superconducting
quantum interference devices and nano-superconducting transistors \cite{Franceschi-Nature5,Huefner-PRB79}.
The possibility of inducing superconductivity in graphene was explored by depositing metal atoms on it.
The superconducting state in metal coated graphene was firstly discussed through a plasmon-mediated mechanism in 2007 \cite{Uchoa-PRL98}.
Subsequent DFT calculations suggested that superconducting transition temperatures ($T_c$) are equal to 8.1 K for monolayer LiC$_6$,
and 1.4 K for monolayer CaC$_6$ \cite{Profeta-Nature8}.
Lately, superconductivity at 7.4 K in Li-intercalated few-layer graphene \cite{Tiwari-arXiv},
and 6 K in Ca-decorated graphene \cite{Chapman-arXiv} has been realized in experiment.
Similar calculations were also carried out for silicene and phosphorene. The $T_c$s were found to be about
15.5 K and 12.2 K for electron-doped silicene and phosphorene under certain tensile strain \cite{Wan-EPL104,Shao-EPL108,Ge-NJP17}.
It is noteworthy that graphene, silicene, and phosphorene, are either semimetal or semicoductor with
vanished density of states at the Fermi level. A prerequisite to induce superconductivity in these compounds is to introduce charge carriers by doping.

Distinct from above semimetals or semiconductor, $\beta_{12}$ and $\chi_3$ borophenes are inherently metallic \cite{Mannix-Science,Feng-arXiv1}.
So superconductivity in borophenes are expected to be an important and interesting issue and deserve to be investigated.
Herein, we chose $\beta_{12}$ and $\chi_3$ borophenes as two potential candidates to investigate their electron-phonon coupling (EPC) properties and possible superconductivity, based on the first-principles calculations and Eliashberg equations.
We found the bond-stretching $A_g$ mode strongly couple with electrons in $\beta_{12}$ borophene.
In $\chi_3$ borophene, several in-plane phonon modes have large contributions to EPC, especially the $B_{1g}$ mode at $\Gamma$ point.
To accurately determine the EPC constants, we employed recently developed Wannier interpolation technique, which has been implemented in the EPW code (see the Method section for details).
The EPC constants for $\beta_{12}$ and $\chi_3$ borophenes are 0.89 and 0.95, which can give rise to 18.7 K and 24.7 K superconductivity, respectively.

\section{Methods}

In our calculations the plane wave basis method was used \cite{pwscf}. We adopted the generalized gradient
approximation (GGA) of Perdew-Burke-Ernzerhof \cite{pbe} for the exchange-correlation potentials. The norm-conserving pseudopotentials \cite{Troullier-PRB43} were used to model the electron-ion interactions. For the slab model, a vacuum layer with 15 {\AA} in thickness was added to avoid the non-physical coupling between adjacent boron sheets along $c$ axis.
After full convergence test, the kinetic energy cut-off and the charge density cut-off of the plane wave basis were chosen to be 60 Ry and 240 Ry, respectively.
The self-consistent electron densities in $\beta_{12}$ and $\chi_3$ borophenes were evaluated on $60\times40\times1$ and $48\times48\times1$ {\bf k}-point grids,
respectively, combining with Marzari-Vanderbilt smearing technique \cite{Marzari-PRL82_3296} of width 0.02 Ry.
The lattice constants after full relaxation were adopted. The effect of electron doping was simulated
by adding extra electrons. A compensating jellium background
was introduced to avoid numerical divergence in a periodic calculation.

The dynamical matrices and phonon perturbation potentials \cite{Giustino-PRB76} were calculated on a $\Gamma$-centered 12$\times$8$\times$1 mesh for $\beta_{12}$ borophene and a 12$\times$12$\times$1 mesh for $\chi_3$ borophene, within the framework of density-functional perturbation theory \cite{Baroni-RMP73_515}. Maximally localized Wannier functions (MLWFs) \cite{Mostofi-CPC178} were determined on the same meshes of the Brillouin zone as that used in above phonon calculations.
Random projection method, including nine (seven) Wannier functions, was used to construct MLWFs in $\beta_{12}$ ($\chi_3$) borophene.
Eventually, fine electron (phonon) grids of $960\times640\times1$ ($240\times160\times1$) and $720\times720\times1$ ($180\times180\times1$) were used to interpolate the EPC quantities in $\beta_{12}$ and $\chi_3$ borophenes with the EPW code \cite{Noffsinger-CPC181}.
In $\beta_{12}$ borophene, Dirac $\delta$-functions for electrons and phonons are replaced by smearing functions with widths of 15 and 0.2 meV, respectively.
These two quantities were selected to be 35 and 0.2 meV in $\chi_3$ borophene. The convergence test of EPC constant is given in the Appendix A.

EPC was calculated based on Eliashberg equations \cite{Eliashberg-ZETF38_966}. The Eliashberg spectral function reads
\begin{equation}
\label{eq:spectral}
\alpha^2F(\omega)=\frac{1}{2\pi N(e_F)}\sum_{{\bf q}\nu}\delta(\omega-\omega_{{\bf q}\nu})\frac{\gamma_{{\bf q}\nu}}{\hbar\omega_{{\bf q}\nu}},
\end{equation}
in which $\omega_{{\bf q}\nu}$ and $\gamma_{{\bf q}\nu}$ are the frequency and linewidth for phonon mode $\nu$ at wavevector ${\bf q}$,
$N(e_F)$ is the density of states at the Fermi level.
The total EPC constant for investigated compound can be determined by either Brillouin-zone summation or the frequency-space integration.
\begin{equation}
\label{eq:lambda}
\lambda=\sum_{{\bf q}\nu}\lambda_{{\bf q}\nu}=2\int\frac{\alpha^2F(\omega)}{\omega}d\omega,
\end{equation}
where $\lambda_{{\bf q}\nu}$ is EPC constant for phonon mode ${\bf q}\nu$.

Using the McMillian-Allen-Dynes formula \cite{Allen-PRB6_2577,Allen-RPB12_905}, we can evaluate the superconducting transition temperature as follows.
\begin{equation}
\label{eq:Tc}
T_c=\frac{\omega_{\text{log}}}{1.2}\exp\Big[\frac{-1.04(1+\lambda)}{\lambda(1-0.62\mu^*)-\mu^*}\Big],
\end{equation}
where $\mu^*$ is the effective screened Coulomb repulsion constant whose value is generally chosen to be
between 0.1 and 0.15 \cite{Richardson-PRL78_118,Lee-PRB52_1425}, and $\omega_{\text{log}}$ is defined through 
\begin{equation}
\label{eq:omega}
\omega_{\text{log}}=\exp\Big[\frac{2}{\lambda}\int\frac{d\omega}{\omega}\alpha^2F(\omega)\log\omega\Big].
\end{equation}

\section{Results}

\begin{figure}[thb]
\includegraphics[width=8.6cm]{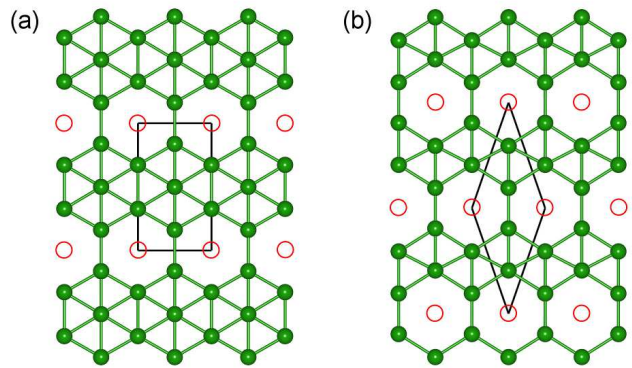}
\caption{(Color online) Top views of borophenes. (a) $\beta_{12}$ borophene.
(b) $\chi_3$ borophene. The green balls and red hollow circles represent the boron atoms and vacancy sites, respectively.
The rectangle and rhombus enclosed by solid black lines denote the unit cells for $\beta_{12}$ and $\chi_3$ borophenes.}
\label{fig:Structure}
\end{figure}

The structures of borophenes observed in experiment were schematically shown in Fig.~\ref{fig:Structure}.
These two structures of borophenes are planar without out-of-plane buckling.
Vacancy sites in them form rectangular and rhombic patterns.
The vacancy concentration and coordination number are two quantities to describe the boron sheets from
global and local points of view, respectively \cite{Tang-PRL99,Wu-ACSNano6}.
The vacancy concentration $\eta$ is defined as ratio between number of vacancy sites and total sites (including vacancy)
in the unit cell. For example, $\eta$ is equal to 1/6 in $\beta_{12}$ borophene, and 1/5 in $\chi_3$ borophene.
The common values of coordination number for 2D boron sheet are four, five, or six.
The ratios among above mentioned three kinds of coordination atoms are 2:2:1 in $\beta_{12}$ borophene,
and 2:2:0 in $\chi_3$ borophene.
After structural optimization, the lattice constants along the $a$ and $c$ axes for $\beta_{12}$ borophene are 2.9346 {\AA} and 5.0840 {\AA},
consistent with previous results \cite{Feng-arXiv1}. In $\chi_3$ borophene, the cell parameter is 4.4566 {\AA},
with the acute angle in the rhombic cell being 38.207$^\circ$.

\begin{figure}[tbh]
\begin{center}
\includegraphics[width=8.6cm]{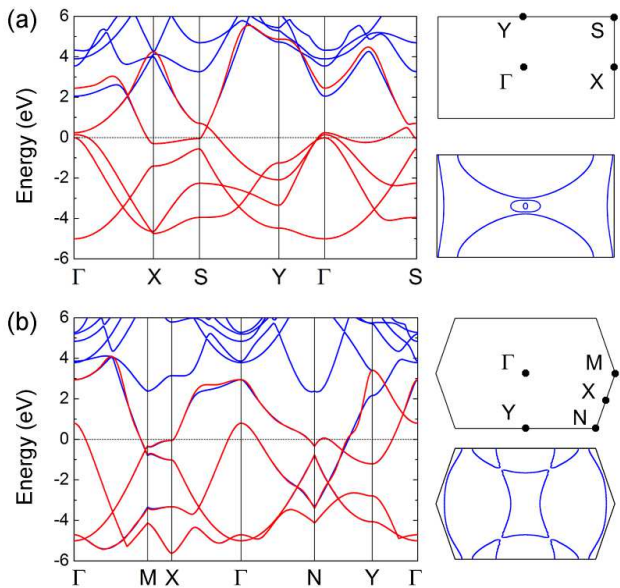}
\caption{(Color online) Calculated electronic structures of $\beta_{12}$ and $\chi_3$ borophenes.
(a) The band structure of $\beta_{12}$ (left panel), the Brillouin zone and the Fermi surfaces (right panel).
(b) The band structure of $\chi_{3}$ (left panel), the Brillouin zone and the Fermi surfaces (right panel).
The blue lines represent band structures given by first-principles calculation.
The red curves are obtained by interpolation of maximally localized Wannier functions (MLWFs).
The Fermi energy is set to zero. The high-symmetry points used in the band-structure calculations are
labeled in the corresponding Brillouin zone.}
\label{fig:NM-Band}
\end{center}
\end{figure}

Firstly, we studied the electronic structures of $\beta_{12}$ and $\chi_3$ borophenes,
using optimized crystal parameters. The calculated band structures and Fermi surfaces were given in Fig.~\ref{fig:NM-Band}.
For both compounds, the band structures unambiguously manifest metallicity, in good agreement with significant
density of states around the Fermi level measured by scanning tunneling spectroscopy \cite{Feng-arXiv1}.
In $\beta_{12}$ borophene, there are three bands crossing the Fermi level.
The first two bands contribute two elliptical hole Fermi sheets surrounding the zone center, as shown on right panel in Fig.~\ref{fig:NM-Band}(a).
Dissimilarly, the third band form two semielliptic electron-like Fermi sheets around the Brillouin zone boundary.
For $\chi_3$ borophene, we found only two bands are partially occupied in the band structure [left panel of Fig.~\ref{fig:NM-Band}(b)].
The lower band in energy give a twisted quadrilateral hole Fermi sheet around $\Gamma$ point.
All other pieces of Fermi sheets in the Fermi surface plot [right panel of Fig.~\ref{fig:NM-Band}(b)], are electron Fermi sheets, originated from the second partially occupied band.
The Wannier interpolated band structures show excellent agreement with these obtained by first-principles calculations below
2 eV. This forms a solid foundation for subsequent EPW calculation.

\begin{figure}[tbh]
\begin{center}
\includegraphics[width=8.6cm]{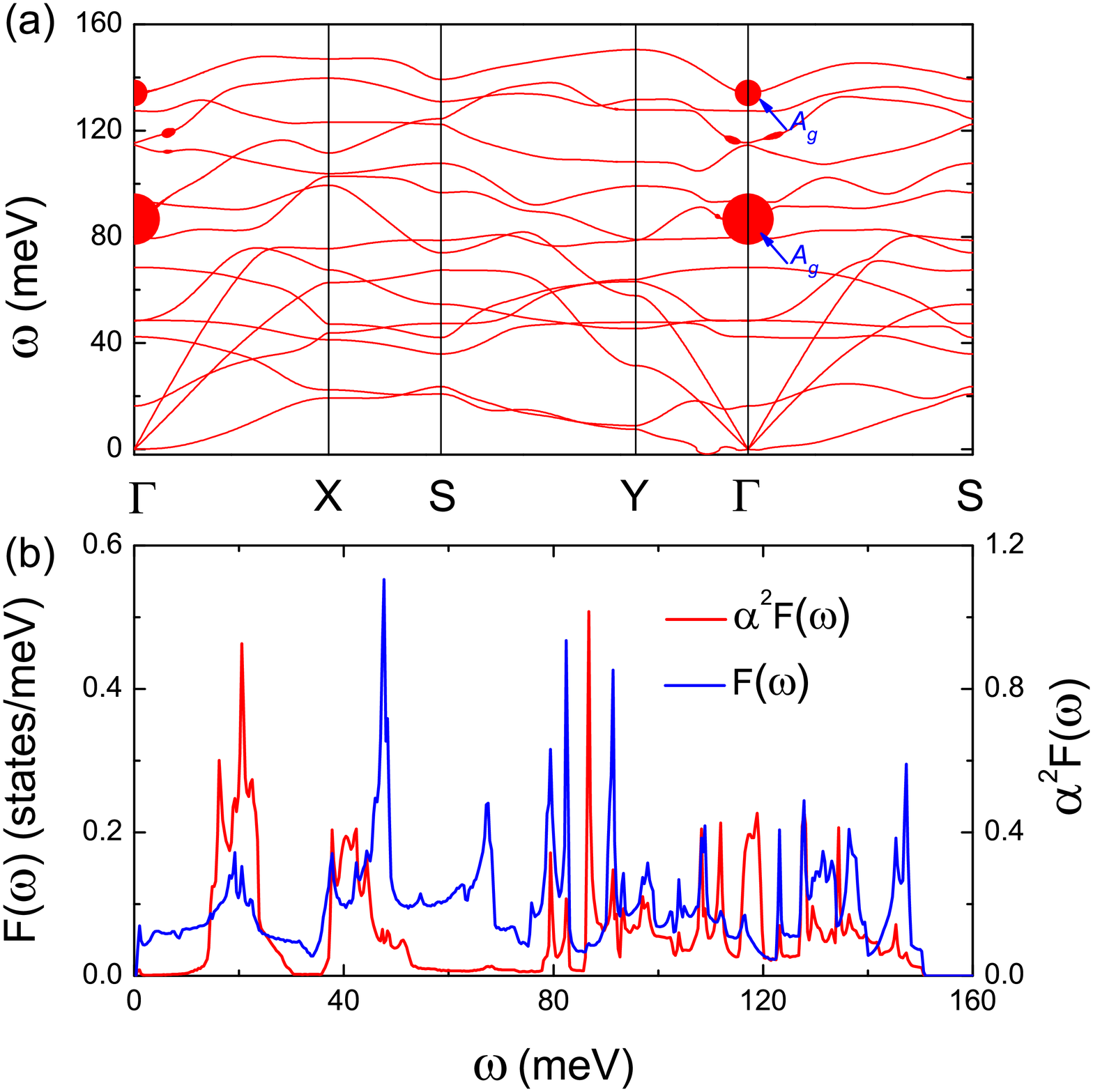}
\includegraphics[width=8.6cm]{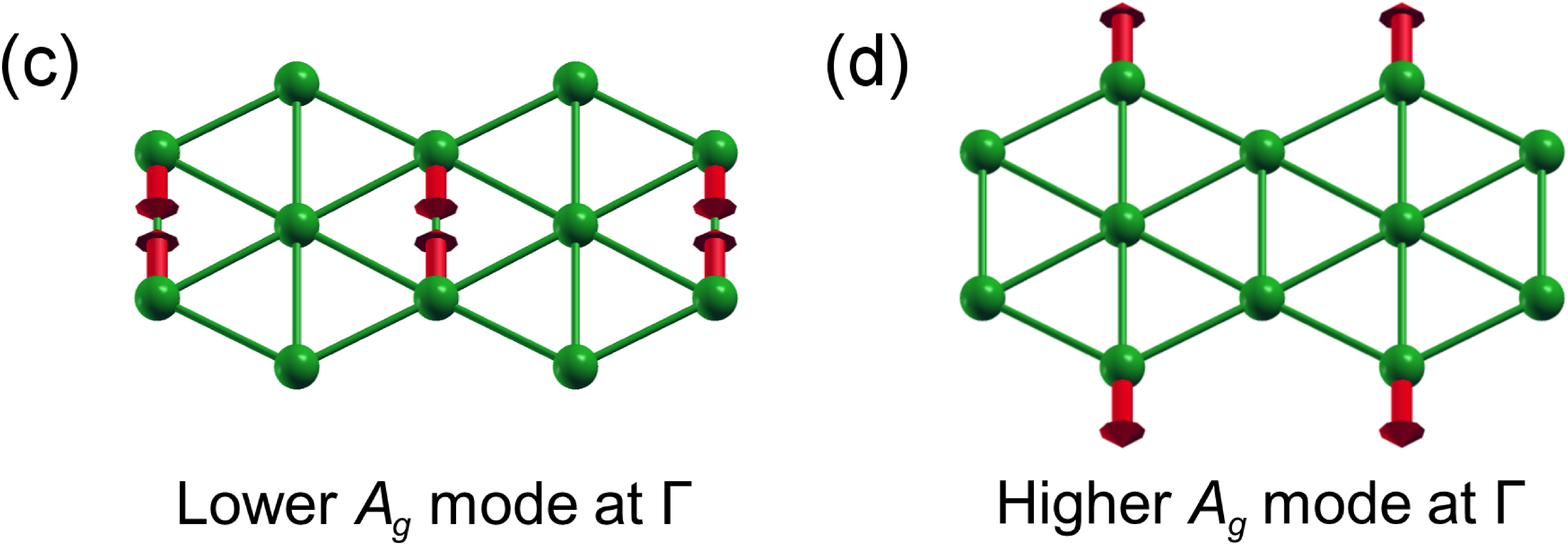}
\caption{(Color online) Lattice dynamics of $\beta_{12}$ borophene. (a) Phonon band structures, in
which the linewidth for phonon mode $\bf{q}\nu$ (i.e., $\gamma_{\bf{q}\nu}$) are represented
by the widths of red lines. (b) Phonon density of states (DOS) $F(\omega)$ and Eliashberg spectral function $\alpha^2F(\omega)$.
(c)-(d) Vibrational pattern for two $A_g$ phonon modes at $\Gamma$.
The red arrows and their lengths denote the directions and relative
amplitudes of these vibrational modes, respectively. }
\label{fig:Phonon-B}
\end{center}
\end{figure}

Secondly, to explore the possible superconductivity in borophene, we calculated
the phonon spectra, Eliashberg spectral functions. The obtained results
for $\beta_{12}$ borophene were shown in Fig.~\ref{fig:Phonon-B}.
Almost all the phonon modes have positive frequencies, except for the transverse branch near $\Gamma$ point with
maximum negative value being -1.75 meV [Fig.~\ref{fig:Phonon-B}(a)].
The main peak in phonon density of sates (DOS) $F(\omega)$ locates
at about 50 meV [Fig.~\ref{fig:Phonon-B}(b)], due to the dispersionless nature of phonon bands around this frequency.
For EPC in $\beta_{12}$ borophene, we found there are two $A_g$ phonon modes at $\Gamma$ point, which have large coupling with electrons,
and the linewidth of lower $A_g$ mode is larger.
The vibrational patterns in real space for these two $A_g$ phonon modes were schematically shown in Fig.~\ref{fig:Phonon-B}(c)-(d), respectively.
Both $A_g$ phonon modes correspond to in-plane bond-stretching mode.
It is also interesting that the movements of six-coordination atoms in the two $A_g$ phonon modes are negligible.
Eliashberg spectral function $\alpha^2F(\omega)$ is a central quantity, through which we can determine the EPC constant $\lambda$ [Eq.~\eqref{eq:lambda}] and the logarithmic average frequency $\omega_{\text{log}}$ [Eq.~\eqref{eq:omega}].
The sharp peak in $\alpha^2F(\omega)$ around 87 meV resulted from the contribution of lower $A_g$ mode.
In the calculation of $\alpha^2F(\omega)$, $\gamma_{{\bf q}\nu}$ is divided by the phonon frequency $\omega_{{\bf q}\nu}$ [Eq.~\eqref{eq:spectral}].
And for $\beta_{12}$ borophene, there exist certain low-frequency phonon modes whose $\gamma_{{\bf q}\nu}$/$\omega_{{\bf q}\nu}$ is larger than that of $A_g$ mode.
The emergence of a peak in Elishberg spectral function around
20 meV is inevitable. We found $\lambda$=0.89, $\omega_{\text{log}}$=27.87 meV, and $T_c$=18.7 K (setting $\mu^*$ to 0.1) for $\beta_{12}$ borophene.

\begin{figure}[tbh]
\includegraphics[width=8.6cm]{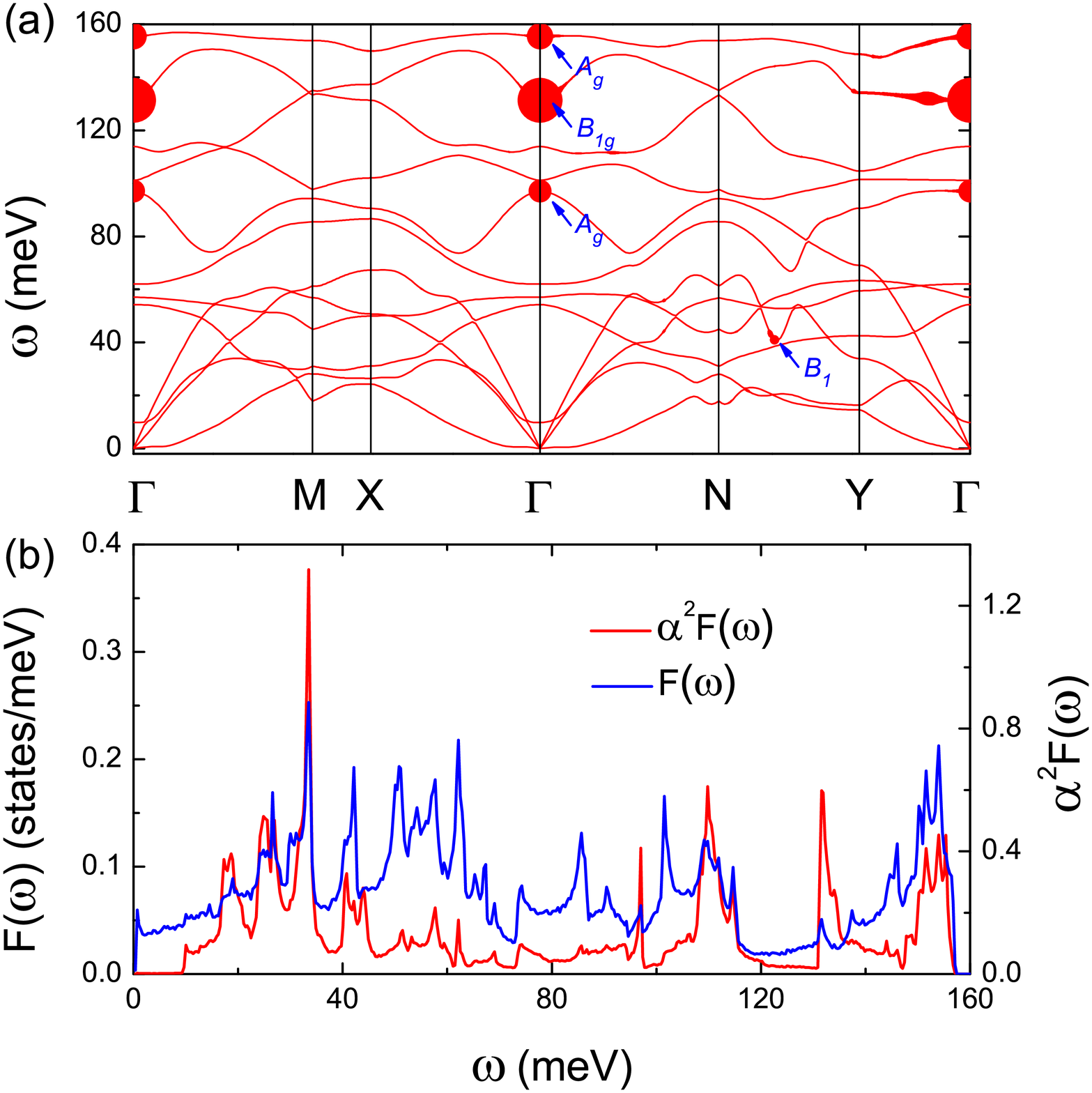}
\includegraphics[width=8.6cm]{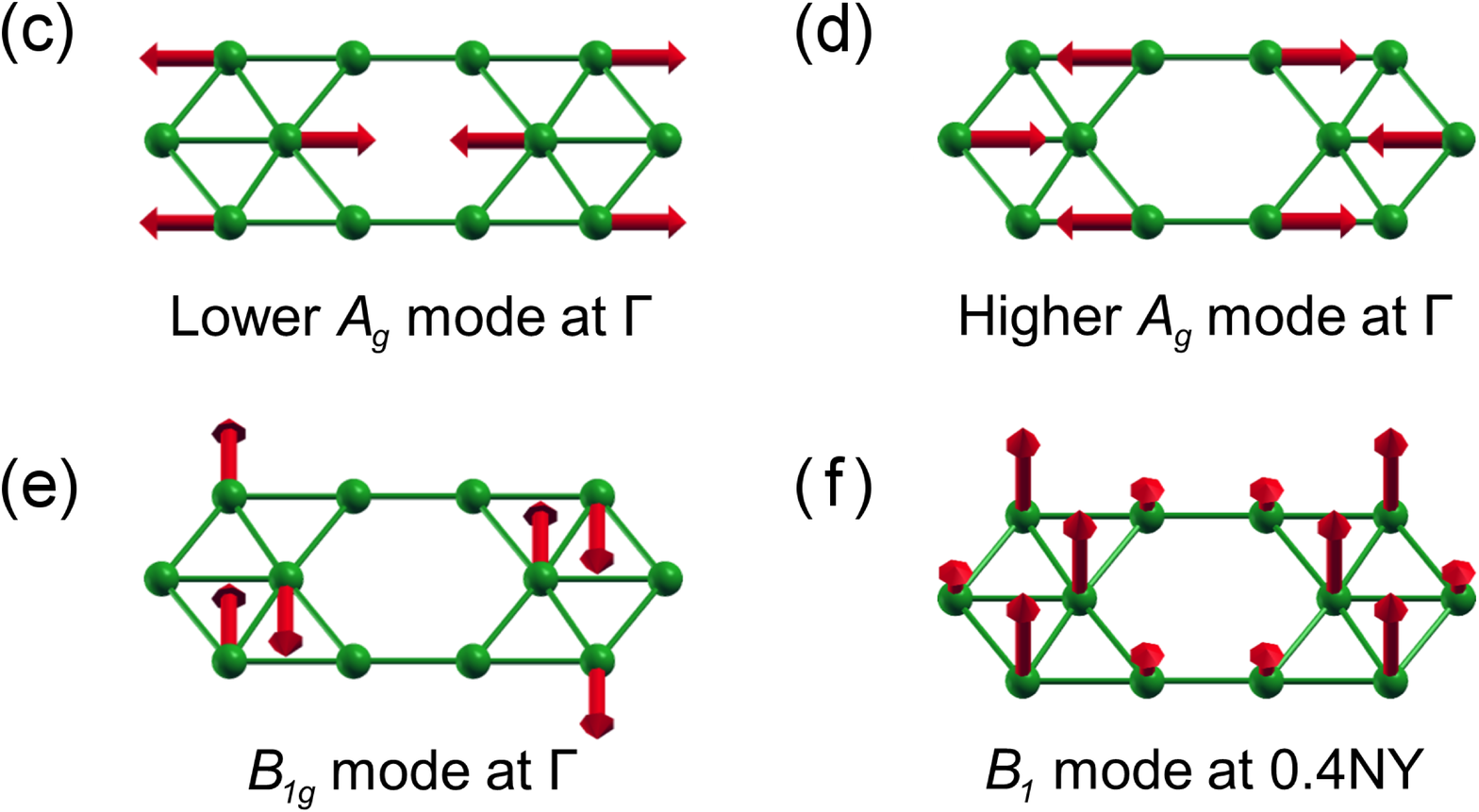}
\caption{(Color online) Lattice dynamics of $\chi_3$ borophene. (a) Phonon band structures, in
which the linewidth for phonon mode $\bf{q}\nu$ (i.e., $\gamma_{\bf{q}\nu}$) are represented
by the widths of red lines. (b) Phonon DOS $F(\omega)$ and Eliashberg spectral function $\alpha^2F(\omega)$.
(c)-(f) Vibrational patterns for important phonon modes.
The red arrows and their lengths denote the directions and relative amplitudes of these vibrational modes, respectively.}
\label{fig:Phonon-C}
\end{figure}

The lattice dynamics and EPC properties of $\chi_3$ borophene were presented in Fig.~\ref{fig:Phonon-C}.
Due to the translation invariant symmetry,
the frequencies of three acoustic phonons vanish at $\Gamma$ point, as they should be [Fig.~\ref{fig:Phonon-C}(a)].
A small negative phonon frequency about -0.25 meV was also found for the transverse phonon branch near $\Gamma$ point.
There is a broadening peak in $\alpha^2F(\omega)$, extending from 15 meV to 40 meV [Fig.~\ref{fig:Phonon-C}(b)].
The fine structures in this peak are resulted from $F(\omega)$, due to the Dirac $\delta$ function in Eq.~\eqref{eq:spectral}.
The Eliashberg spectral function resembles the phonon DOS above 40 meV.
In $\chi_3$ borophene, we identified four phonon modes with large EPCs.
To be specific, they are $B_{1g}$ phonon mode, two $A_g$ phonon mode at $\Gamma$ point, and $B_1$ phonon mode at about 0.4$NY$.
The above mentioned zone-centered phonon modes involve the in-plane atomic movements.
While the $B_1$ mode correspond to out-of-plane displacements of boron atoms.
The $B_{1g}$ phonon mode at $\Gamma$ has the largest linewidth, which give rise to a peak in $\alpha^2F(\omega)$ around 135 meV.
Finally, the calculated $\lambda$, $\omega_{\text{log}}$, and $T_c$ are 0.95, 33.10 meV, and 24.7 K, respectively.
The EPC constant in MgB$_2$ given by Wannier interpolation technique is about 0.74 \cite{Calandra-PRB82,Margine-PRB87}.
Thus both borophenes possess larger EPC constants in comparison with MgB$_2$.

The dynamical stability of a crystal structure is reflected by none imaginary phonon frequency in the Brillouin zone.
But in both borophenes, our calculation showed the frequency of transverse branch become imaginary near $\Gamma$ point.
For 2D compounds, two acoustical branches are linear with {\bf q} near $\Gamma$ point.
On the contrary, due to the rapid decay of force constants related to the transverse movements of atoms, the transverse branch
always show a quadratic dispersion near $\Gamma$ point \cite{Liu-PRB76}. The imaginary phonon frequency of transverse branch
near $\Gamma$ points were also found in the simulations of germanene \cite{Cahangirov-PRL102}, buckled arsenene \cite{Kamal-PRB91},
and other binary monolayer honeycomb structures \cite{Sahin-PRB80}. The occurrence of imaginary frequency was
attributed to the numerical difficulties in accurate calculation of rapid decaying interatomic forces, and is not
a sign of structural transition \cite{Sahin-PRB80}. Moreover, these imaginary frequencies do not affect our EPC results.

\begin{figure*}[tbh]
\includegraphics[width=17.2cm]{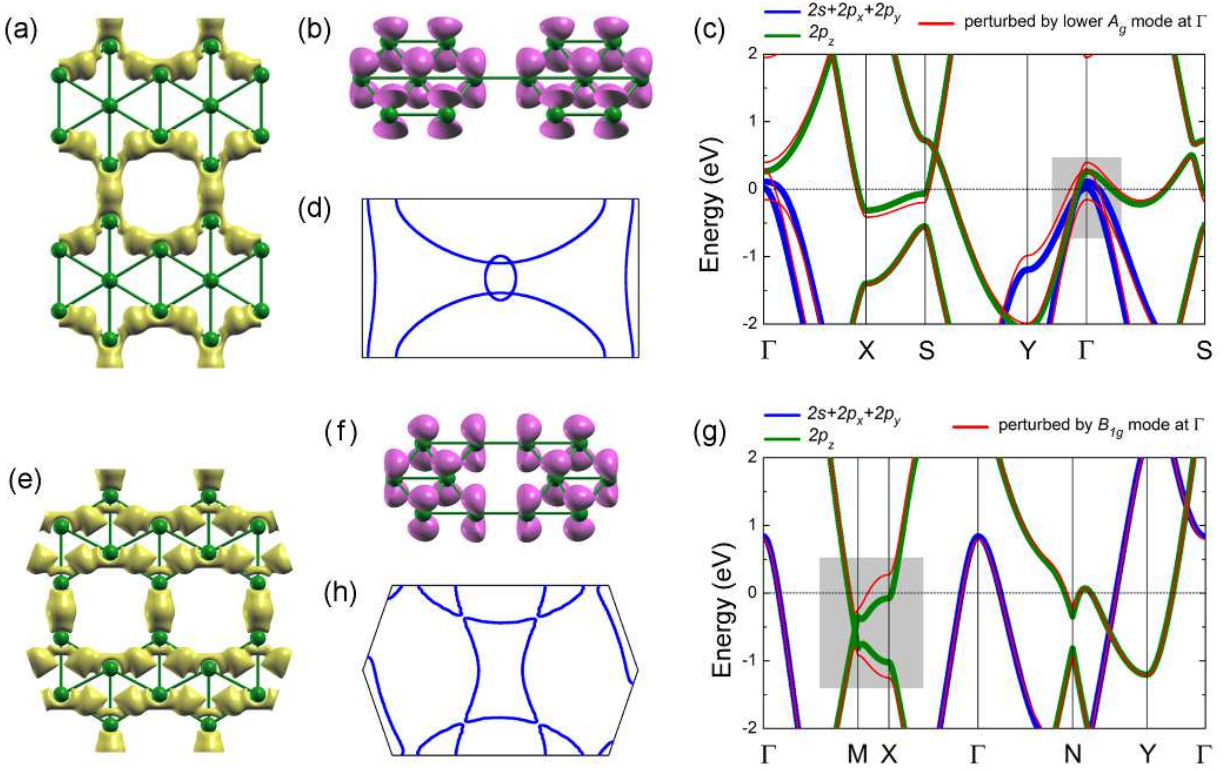}
\caption{(Color online)
Positive DCD (isovalue: 2.8$\times 10^{-2}$ $e$/bohr$^3$) (a), negative DCD (isovalue: -8.0$\times 10^{-3}$ $e$/bohr$^3$) (b),
orbital-resolved and phonon perturbed band structures (c), and Fermi surfaces perturbed by $B_{3g}$ phonon mode (d) in $\beta_{12}$ borophene.
Positive DCD (isovalue: 2.45$\times 10^{-2}$ $e$/bohr$^3$) (e), negative DCD (isovalue: -8.0$\times 10^{-3}$ $e$/bohr$^3$) (f),
orbital-resolved and phonon perturbed band structures (g), and Fermi surfaces perturbed by $B_{3g}$ phonon mode (h) in $\chi_{3}$ borophene.
The widths of blue and green lines in the band structures stand for the weights of $2s+2p_x+2p_y$ and $2p_z$ orbitals of boron in given Kohn-Sham state.
The red lines represent the band structures perturbed by $B_{3g}$ phonon modes at $\Gamma$ point.
The obvious deviations in band structures before and after phonon perturbation are enclosed by gray shadows.}
\label{fig:charge}
\end{figure*}

In order to have an insight into the physical origin of strong coupling between electrons and aforementioned phonons,
we calculated differential charge densities (DCD), i.e., charge density minus superposition of atomic charge densities, to
determine the bonding nature in borophenes [see Fig.~\ref{fig:charge}].
For $\beta_{12}$ borophene, the positive part of DCD mostly distribute in the middle region of neighboring
four-coordination and five-coordination boron atoms [Fig.~\ref{fig:charge}(a)].
This unambiguously shows the formation of in-plane $sp^2$-hybridization-like $\sigma$ bonds between these two kinds of boron atoms.
The negative isosurface of DCD locate on the top of each boron atom, resembling the B-$2p_z$ orbital.
Boron has three valance electrons. For an isolated boron atom, two electrons with opposite spin orientations will firstly occupy the B-$2s$ orbital.
Due to the degeneracy among B-$2p_x$, B-$2p_y$, and B-$2p_z$, the third valance electron has the same probability to take one of the three orbitals.
But in borophene crystal, the energies of $sp^2$-hybridized orbitals are evidently lower than that of B-$2p_z$ orbital, the three valance electrons trend to
fill the $sp^2$-hybridized bonding states, as a result the occupation number in B-$2p_z$ orbital is significantly reduced.
This is the reason for $p_z$-orbital-like negative DCD shown in Fig.~\ref{fig:charge}(b).
The B-$2p_z$ orbital will overlap with its neighboring ones, forming $\pi$ bands.
The DCD and bonding characteristics in $\chi_3$ borophene exhibit similar behavior as that in the $\beta_{12}$ one [Fig.~\ref{fig:charge}(e) and (f)].

We then projected the Kohn-Sham state $\Psi_{n\bf{k}}(r)$ to boron atomic orbitals. Fig.~\ref{fig:charge}(c) and (g) show the calculated orbital-resolved
band structures for $\beta_{12}$ and $\chi_3$ borophenes, respectively. The widths of blue and green lines denote the contributions of $2s+2p_x+2p_y$ and $2p_z$ orbitals to given Kohn-Sham states. Detail analysis shows the uniform width of blue (green) lines is due to the fact that weights projected onto $2s+2p_x+2p_y$ ($2p_z$) are close to 1.0, suggesting
the blue (green) lines are pure $\sigma$ ($\pi$) bands. According to orbital-resolved band structures, we can further infer
the two hole Fermi sheets surrounding $\Gamma$ point in $\beta_{12}$ borophene [Fig.~\ref{fig:NM-Band}(a)] are from $\sigma$-bonding bands.
For $\chi_3$ borophene, top and bottom pieces of $\Gamma$-centered twisted quadrilateral Fermi sheets [Fig.~\ref{fig:NM-Band}(b)] stem from $\pi$ bands,
$\sigma$-bonding bands contribute to the left and right pieces. Meanwhile, the first Fermi sheet from Y point to N point in the Brillouin zone
also belong to $\sigma$-bonding bands.

Since the lower $A_g$ phonon mode and the $B_{1g}$ mode at $\Gamma$ have the largest phonon linewidth in
$\beta_{12}$ and $\chi_3$ borophenes, respectively.
It is important to answer which bands or Fermi surfaces couple to $A_g$ or $B_{1g}$ phonons.
A simple method to investigate EPC for a $\Gamma$-point optical phonon mode is to calculate the deformation potential caused by this phonon displacement \cite{Khan-PRB29,An-PRL86}.
We moved the atomic positions according to the $A_g$ or $B_{1g}$ phonon displacement, and recalculated the band structures perturbed by phonon, namely deformation potential.
In the calculation of perturbed band structures and Fermi surfaces, the $A_g$ and $B_{1g}$ phonon displacements are normalized to 0.05 {\AA} to obtain a clear vision. Fig.~\ref{fig:charge}(c) and (g) show the band structures before and after phonon perturbation.
For $\beta_{12}$ borophene, the $\sigma$-bonding bands near $\Gamma$ are dramatically changed with respect to the equilibrium band structures [see gray shadow in Fig.~\ref{fig:charge}(c)].
On the contrary, all the $\pi$ bands are almost unaffected. The perturbed Fermi surfaces again confirm above effects, with obvious distortions of the two $\Gamma$-centered hole Fermi sheets [Fig.~\ref{fig:charge}(d)].
This means these two hole Fermi sheets, originated from $\sigma$-bonding bands, strongly couple to $A_{g}$ phonon in $\beta_{12}$ borophene.
For $\chi_3$ borophene, the positions of $\pi$ band along $M$-$X$ line are shifted after phonon perturbation [Fig.~\ref{fig:charge}(g)],
suggesting strong coupling between these bands and $B_{1g}$ phonon mode.
As a consequence, the Fermi sheets near lower right and top left corners considerably shrink [Fig.~\ref{fig:charge}(h)].

\section{Discussion}

Among the theoretically proposed models for boron sheet, neither $\beta_{12}$ nor $\chi_3$
is the ground-state structure \cite{Wu-ACSNano6}. Thus, the interfacial boron-silver interaction should play an important role to stabilize these two monolayer structures on Ag(111) substrate. Experiment indeed observed a charge transfer from Ag(111) to boron sheet \cite{Feng-arXiv1}. The slab simulation
identified a tiny charge transfer ($\sim$0.03 $e$/boron atom) from the Ag substrate to borophene \cite{Feng-arXiv1}.
So borophenes are slightly electron doped.
After deposition, the periodicity of borophene observed in experiment along the [$\bar{1}\bar{1}2$] direction of the Ag(111) surface is 15.0688 {\AA} \cite{Feng-arXiv1}.
Due to lattice mismatch, this will impose 2.7\% and 3.3\% uniaxial tensile strains on $\beta_{12}$ and
$\chi_3$ borophenes, respectively. For electron doping, our calculations showed the superconducting transition temperatures are slightly suppressed to 12.9 K and 21.6 K in these two compounds. After taking the tensile strains into consideration, the $T_c$s are further reduced by about 57\% and 41\% in $\beta_{12}$ and
$\chi_3$ borophenes, with respect to the free standing case [see Table I].

The strong EPC may rise the question about lattice instability.
We found the frequency of lower $A_{g}$ ($B_{1g}$) phonon change from 87.4 (131.3) meV to 88.5 (118.4) meV in $\beta_{12}$ ($\chi_3$) borophene,
by simulating the combined effects of charge transfer (0.03 $e$/boron) and tensile strain (2.7\% in $\beta_{12}$, 3.3\% in $\chi_3$) provided by the Ag substrate.
Thus no obvious phonon soften is found, when taking the substrate effect into consideration.
Especially, the slightly hardened $A_{g}$ phonon indicates the substrate can further stabilize $\beta_{12}$ borophene at the cost of degressive $T_c$.
The strength of EPC is reserved to some extent in both borophenes, whose $T_c$s still above the predicted 8.1 K superconductivity in graphene.
On the other hand, the crystal structure of deposited boron sheet strongly depends on the nobel-metal substrate \cite{Zhang-Angew127}.
There is a probability to grow a novel 2D boron sheet, whose $T_c$ is higher than that of currently investigated,
by adopting different substrate.

We have discussed the influences of tensile strain and electron doping on superconductivity in suspended borophenes,
without including Ag substrate in the calculation. But the screening effect of Ag substrate is also very important. Herein, we constructed a minimal model of $\chi_3$ borophene on Ag(111) to estimate the effect of Ag substrate. Considering the tremendous workload in EPC computing, we adopted a single silver atom layer to represent the Ag(111) surface.
This model has a 3$\times$3-Ag(111)-surface unit cell, in which there are 9 Ag atoms and 20 boron atoms. We labelled above model as $\chi_3$/SL-Ag(111).
The EPC of $\chi_3$/SL-Ag(111) was also calculated by Wannier interpolation technique (see Appendix B for detail).
We found that $T_c$ of $\chi_3$/SL-Ag(111) is 10.0 K, which is close to 14.5 K given in Table I.
Although the $T_c$ of $\chi_3$ borophene is markedly reduced by Ag substrate, it is still higher than that found in superconducting graphene-based compounds.

\begin{table}
\caption{Superconductivity in 2D compounds, including graphene \cite{Profeta-Nature8},
silicene \cite{Wan-EPL104}, phosphorene \cite{Shao-EPL108}, stanene \cite{Shaidu-arXiv}, and borophene, predicted by first-principles calculations.
Additional charge carriers were introduced by either deposition of guest atoms or electron doping. The strain is define
as $\varepsilon=(a-a_0)/a_0\times100\%$, in which $a_0$ and $a$ are the lattice constants of equilibrium and strained structures, respectively. }
\label{table:energy}
\begin{tabular}{lccccc}
  \hline
  compounds & doping & $\varepsilon$ (\%) & $\lambda$ & $\omega_{\text{log}}$ (meV) & $T_c$ (K) \\
  \cline{1-6}
  graphene & Li deposition & 0.0 & 0.61 & 34.44 & 8.1 \\
  silicene & 0.44 $e$/atom & 0.0 & 0.44 & 29.38 & 1.7 \\
           & 0.39 $e$/atom & 5.0 & 1.04 & 26.17 & 15.5 \\
  phosphorene & 0.10 $e$/atom & 0.0 & 0.54 & 21.86 & 4.2 \\
              & 0.10 $e$/atom & 8.0 & 1.31 & 10.64 & 12.2 \\
  stanene  & Li deposition & 0.0 & 0.65 & 5.25 & 1.3 \\
  $\beta_{12}$ borophene & 0.0 & 0.0 & 0.89 & 27.87 & 18.7 \\
                         & 0.03 $e$/atom & 0.0 & 0.72 & 30.00 & 12.9 \\
                         & 0.03 $e$/atom & 2.7 & 0.55 & 39.69 & 8.0 \\
  $\chi_{3}$ borophene & 0.0 & 0.0 & 0.95 & 33.10 & 24.7 \\
                       & 0.03 $e$/atom & 0.0 & 0.89 & 32.04 & 21.6 \\
                       & 0.03 $e$/atom & 3.3 & 0.66 & 41.09 & 14.5 \\
  \hline
\end{tabular}
\end{table}

Currently, several 2D compounds were predicted to be phonon-mediated superconductors in literatures by first-principles simulations.
We listed the obtained quantities of these compounds in Table I to make a comparison. In the table, graphene, silicene, and stanene are Dirac semimetals, phosphorene is a semiconductor. Additional charge carriers must be introduced to achieve superconductivity. For strain-free situation, the EPC constants of borophenes are the largest ones among listed compounds. The $T_c$ of 24.7 K in $\chi_3$ borophenes is the highest one among these five 2D compounds without and with strain.
After applying tensile strain, the superconducting transition temperatures of silicene and phosphorene increase to 15.5 K and 12.2 K, respectively.
The influence of heavy electron/hole doping and/or large tensile/compressive strain on superconductivity in borophenes are also an important research subject, which need to be clarified in further study.

In conclusion, we have studied the electronic structure and EPC in two kinds of borophene.
The superconducting transition temperatures are around 20 K,
which is higher than observed 7.4 K superconductivity in graphene.
General speaking, it is challenging for experiment to confirm superconductivity in 2D compounds on a conductive substrate.
But very recently, four-point probe electrical transport measurements \cite{Ge-Nat} and two-coil mutual inductance
measurements \cite{Zhang-Sci} were successfully used to detect zero resistance and Meissner effect, respectively, in monolayer FeSe film grown on Nb-doped SrTiO$_3$ substrate by molecular beam epitaxy.
Here, Nb-doped SrTiO$_3$ is also a conductive substrate.
Considering borophene has been synthesized in experiment, our prediction can be directly examined by above mentioned two techniques.

\textit{Note added}. After posted on arXiv, we noticed a paper by E. S. Penev, A. Kutana, and B. I. Yakobson,
who utilized local density approximation and norm-conserving pseudopotentials to compute electron-phonon couplings in three structures of borophenes and
drew similar conclusions \cite{Penev-Nano}.

\begin{acknowledgments}
This research is supported by National Natural Science Foundation of China (Nos. 11404383 and 11474004)
and Zhejiang Provincial Natural Science Foundation of China under Grant No. LY17A040005.
This work is also sponsored by K.C.Wong Magna Fund in Ningbo University.

\end{acknowledgments}

\appendix
\section{Convergence test of electron-phonon coupling constant}

\begin{figure}[tbh]
\includegraphics[width=8.6cm]{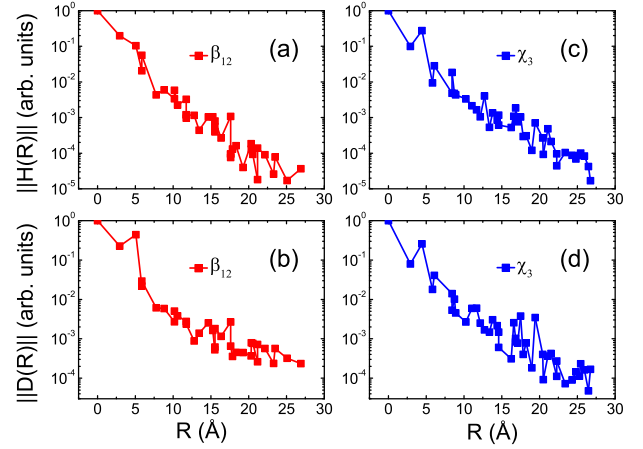}
\caption{(Color online)
Spatial decay of electronic Hamiltonian and dynamical matrix in the Wannier representation.
(a) and (c) represented decay of electronic Hamiltonian for $\beta_{12}$ and $\chi_3$ borophene.
Decay of dynamical matrix for $\beta_{12}$ and $\chi_3$ borophene were shown in (b) and (d),
respectively.}
\label{fig:decay}
\end{figure}

The correctness and accuracy of EPC calculation with Wannier interpolation technique strongly rely on
the spatial localization of the Hamiltonian ($\|H(R)\|$) and dynamical matrix ($\|D(R)\|$) in the
Wannier representation. We have carefully examined the spatial decay of above two quantities,
which are shown in Fig.~\ref{fig:decay}.
The expressions of $\|H(R)\|$ and $\|D(R)\|$ read \cite{Giustino-PRB76},
\begin{equation}
\nonumber
\begin{array}{ll}
\|H(R)\|=\text{max}_{mn,{\bf R_e}-{\bf R_e'}=R}|\langle m{\bf R_e'}|\hat{H}^{el}|n {\bf R_e} \rangle|, \text{and} \\
\|D(R)\|=\text{max}_{\kappa\kappa'\alpha\alpha',|{\bf R}_p-{\bf R}_p'|=R}|\langle\kappa'\alpha'{\bf R}_p|\hat{D}^{ph}|\kappa\alpha{\bf R}_p' \rangle|.
\end{array}
\end{equation}
$\hat{H}^{el}$ and $\hat{D}^{ph}$ are the electron Hamiltonian and phonon dynamical matrix in the Bloch representation, respectively.
$m$, $n$ stand for the index of Wannier functions in one unit cell. ${\bf R_e}$, ${\bf R_e'}$, ${\bf R}_p$, and ${\bf R}_p'$ represent the unit cell positions.
$\kappa$ and $\kappa'$ denote the locations of atoms in one unit cell. Both $\alpha$ and $\alpha'$ traverse all the three different directions of cartesian coordinate.

The coarse {\bf k}-meshes used in the determination of MLWFs were 12$\times$8$\times$1 and 12$\times$12$\times$1 for $\beta_{12}$ and $\chi_3$ borophenes.
This means the borophene crystals we computed are
Wigner-Seitz supercells corresponding to 12$\times$8$\times$1 or 12$\times$12$\times$1 replicas of the primitive cell in periodic boundary simulation.
The hopping distance among two MLWFs can extend to about 17 {\AA} in both borophenes.
While the hopping term of Hamiltonian in the Wannier representation with distance outside the Wigner-Seitz supercell will be truncated in the
generalized Fourier interpolation [see Eq.(31) in Ref. \onlinecite{Giustino-PRB76}].
From Fig.~\ref{fig:decay}(a) and (b), we found the Hamiltonians in Wannier representation ($\|H(R)\|$) show exponential decay with increasing $R$.
Especially, the ratios between $\|H(\sim17\text{{\AA}})\|$ and $\|H(0)\|$ decrease to 1.07$\times$10$^{-3}$ and 7.60$\times$10$^{-4}$ for $\beta_{12}$ and $\chi_3$ borophenes.
The error in constructing electron Hamiltonian for an arbitrary {\bf k} point introduced by Fourier interpolation is only 0.1\%.
On the other hand, the dynamical matrix in the Wannier representation is proportional to the interatomic force constants,
whose spatial decay exhibits similar behavior as that for $\|H(R)\|$ [see Fig.~\ref{fig:decay}(c) and (d)].
Thus electron Hamiltonian and phonon dynamical matrix in the Wannier representation constructed on above used coarse {\bf k}- and {\bf q}-meshes
are sufficient to interpolate EPC properties.

\begin{figure}[tbh]
\includegraphics[width=8.6cm]{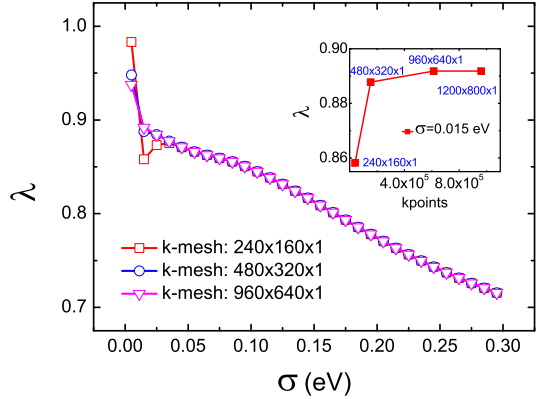}
\caption{(Color online)
Convergence test of EPC constant versus smearing parameter $\sigma$ in $\beta_{12}$ borophene. In the calculation, fine phonon grid was chosen to be 240$\times$160$\times$1. The inset shows the convergence of $\lambda$ for $\sigma$=0.015 eV, with increasing density of {\bf k}-mesh.}
\label{fig:Bconv}
\end{figure}

\begin{figure}[tbh]
\includegraphics[width=8.6cm]{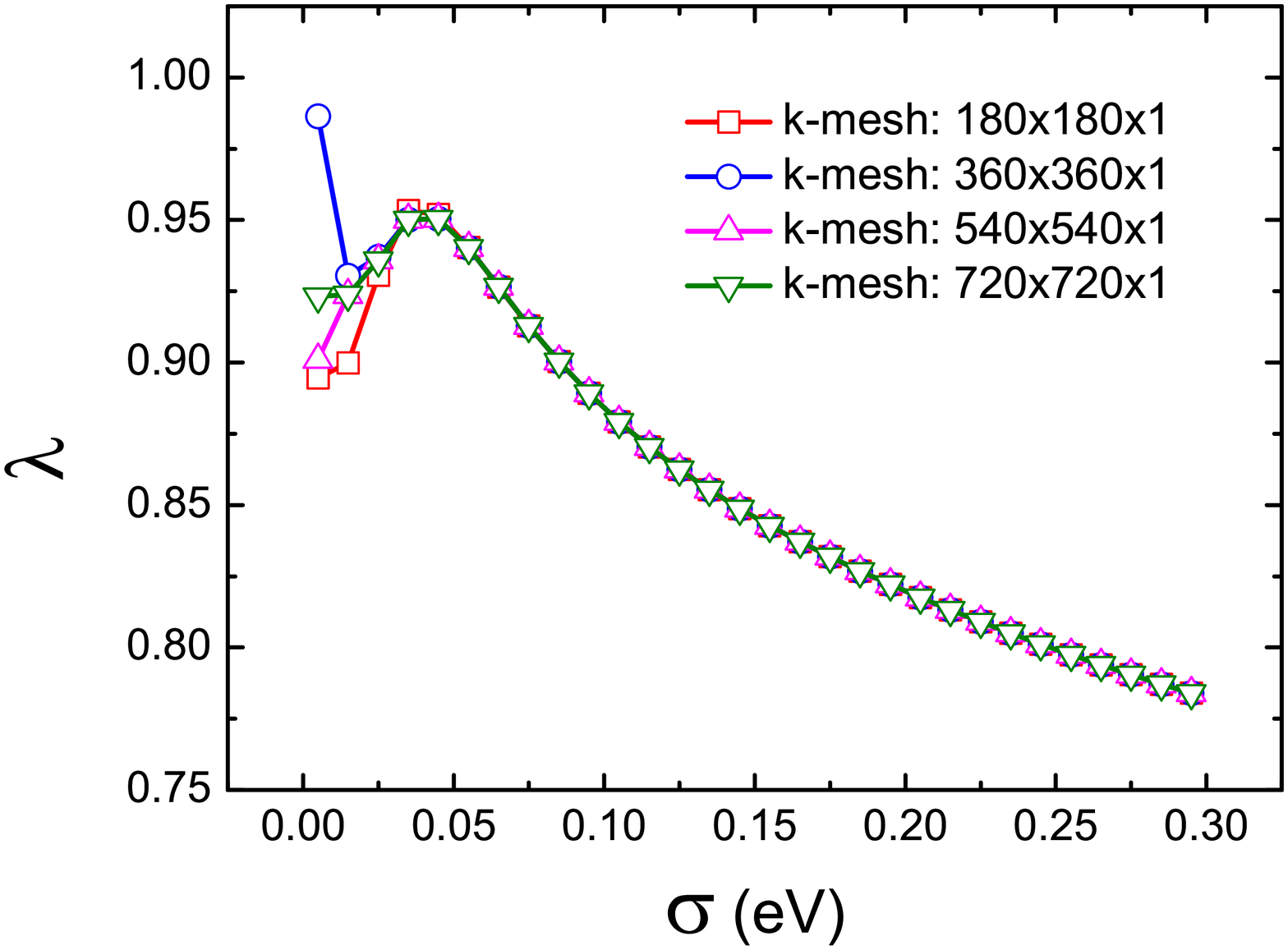}
\caption{(Color online)
Convergence test of EPC constant versus smearing parameter $\sigma$ in $\chi_{3}$ borophene. In the calculation, fine phonon grid was chosen to be 180$\times$180$\times$1.}
\label{fig:Cconv}
\end{figure}

The difficulty to obtain convergent $\lambda$ originate from double $\delta$ functions in calculating the EPC matrix element.
In practical calculation, $\delta$ function is replaced by smearing function with a broadening width $\sigma$.
What we are interested in is the limit that $\sigma
\rightarrow0$ and number of ${\bf k}$ points $\rightarrow\infty$. In order to obtain convergent $\lambda$,
we have extensively tested the convergence versus $\sigma$ and number of ${\bf k}$ points, shown in Fig.~\ref{fig:Bconv} and Fig.~\ref{fig:Cconv}.
Although there is no plateau when $\sigma$ approaching zero in $\beta_{12}$,
the fine meshes 1200$\times$800$\times$1 and 960$\times$640$\times$1 already give convergent $\lambda$ for $\sigma$ being 0.015 eV [see the inset in Fig.~\ref{fig:Bconv}].
Compared with commonly used 0.01$\sim$0.02 Ry for $\sigma$ in the literatures, 0.015 eV is more closer to the limit of $\sigma
\rightarrow0$. Thus we chose EPC properties obtained by {\bf k}-mesh 960$\times$640$\times$1 and $\sigma$=0.015 eV as our results presented in the main text.
For $\chi_3$ borophene, the emergence of a small plateau near 0.035 eV, and the overlapping results among different {\bf k}-meshes at 0.035 eV
clearly indicate that $\lambda$ is convergent for 720$\times$720$\times$1 mesh and $\sigma$=0.035 eV.

The convergence of {\bf q}-mesh for Brillouin zone summation in Eq.~\eqref{eq:lambda} is also tested by enlarging the
fine {\bf q}-meshes to 300$\times$200$\times$1 and 240$\times$240$\times$1 for $\beta_{12}$ and $\chi_3$ borophenes, respectively.
We found the differences in $\lambda$s are merely 2.50$\times$10$^{-4}$ for $\beta_{12}$ and 5.89$\times$10$^{-5}$ for $\chi_3$, with respect to that given in the main text.
We also checked the convergence of phonon frequency on the {\bf k}-mesh that used to determine the self-consistent charge density.
A smearing of 0.02 Ry
together with a mesh of 90$\times$60$\times$1 yielded exactly the same frequencies for all the optical phonon modes at $\Gamma$ point for $\beta_{12}$ borophene.
A similar calculation was also carried out for $\chi_3$ borophene by increasing the {\bf k}-mesh from 48$\times$48$\times$1 to 72$\times$72$\times$1.
And we again did not observe any differences in the frequencies for optical phonon modes at $\Gamma$.

\section{Effect of Ag substrate}

\begin{figure}[tbh]
\begin{center}
\includegraphics[width=8.6cm]{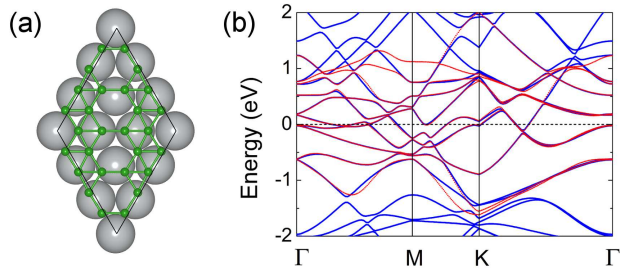}
\caption{(Color online) (a) Ground-state structure of $\chi_3$/SL-Ag(111). The gray balls represent Ag atoms.
The solid lines denote the unit cell.
(b) Band structures of $\chi_3$/SL-Ag(111) obtained by first-principles calculation (blue lines) and interpolation of MLWFs (red lines).}
\label{fig:Ag111}
\end{center}
\end{figure}

\begin{figure}[tbh]
\begin{center}
\includegraphics[width=8.6cm]{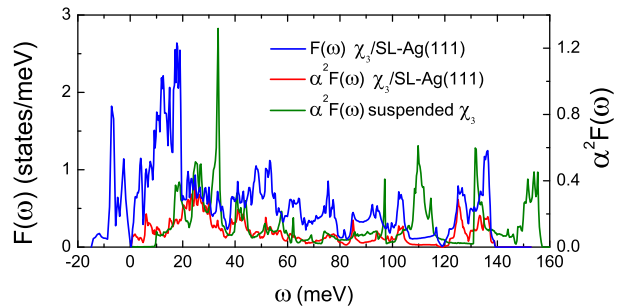}
\caption{(Color online) Phonon DOS and Eliashberg spectral function for $\chi_3$/SL-Ag(111).}
\label{fig:Ag111-PDOS}
\end{center}
\end{figure}

The $\chi_3$/SL-Ag(111) model contains one silver layer and one boron layer in $\chi_3$ configuration.
Currently, we do not investigate $\beta_{12}$/SL-Ag(111) due to the huge number of atoms (105 atoms) in
3$\sqrt{3}\times$5-Ag(111)-surface unit cell \cite{Feng-arXiv1}.
In the calculation of $\chi_3$/SL-Ag(111), 8$\times$8$\times$1 {\bf k}-mesh together with a 0.02 Ry Marzari-Vanderbilt smearing technique
were used to evaluate the self-consistent charge density. The lattice constant was fixed to 8.7 {\AA}, which is
three times as large as that for Ag(111) surface. By relaxation of inner coordinates for silver and boron atoms,
we obtained the ground-state structure of $\chi_3$/SL-Ag(111) [see Fig.~\ref{fig:Ag111}(a)].
Here, we constructed ten MLWFs, with which energy band at arbitrary {\bf k} point can be interpolated.
As can be see in Fig.~\ref{fig:Ag111}(b), the interpolated band structure is in excellent agreement with
the one calculated by first-principles from -0.5 eV to 0.5 eV.

Dynamical matrices and phonon perturbation
potentials were calculated on a coarse 4$\times$4$\times$1 {\bf q}-mesh.
After convergence test, the fine {\bf k}- and {\bf q}-mesh used in EPW code were chosen
to be 240$\times$240$\times$1 and 80$\times$80$\times$1, respectively.
In the phonon DOS [Fig.~\ref{fig:Ag111-PDOS}], the phonon frequencies are significantly softened
after including Ag substrate. For example, the highest phonon frequency is reduced by about 12.5\% with respect to that in suspended $\chi_3$ borophene.
We also found that there are some imaginary phonon frequencies,
which may be related to the fact that only one single Ag layer are contained in $\chi_3$/SL-Ag(111).
The Eliashberg spectral function $\alpha^2F(\omega)$ of $\chi_3$/SL-Ag(111) is red-shifted in comparison with suspended $\chi_3$ borophene [Fig.~\ref{fig:Ag111-PDOS}].
According to Eq.~\eqref{eq:lambda} and Eq.~\eqref{eq:omega}, red-shifted $\alpha^2F(\omega)$ will give rise to
ascendent $\lambda$ and descendent $\omega_{\text{log}}$.
Finally, we found that $\omega_{\text{log}}$=11.48 meV, $\lambda$=1.05, and $T_c$=10.0 K for $\chi_3$/SL-Ag(111).
We also noticed the tensile strain and electron doping can also suppress superconductivity of $\chi_3$ borophene through reducing $\lambda$.
Thus, this indicates that the effect of Ag substrate can not be simply replaced by tensile strain and electron doping.


\end{document}